\pdfoutput=1
\documentclass[aps,pre,twocolumn]{revtex4-1}

\usepackage{amsmath}
\usepackage{amssymb}
\usepackage{graphicx}
\usepackage[margin =1 in]{geometry}
\usepackage{ulem}
\usepackage{mathtools}

\def\be{\begin{equation}}
\def\ee{\end{equation}}
\def\bea{\begin{eqnarray}}
\def\eea{\end{eqnarray}}
\def\bml{\begin{mathletters}}
\def\eml{\end{mathletters}}
\def\bse{\begin{subequations}}
\def\ese{\end{subequations}}
\def\expec{\mathbb{E}}

\def\Var{\text{Var}}

\begin{document}

\title{Stationary moments, diffusion limits, and extinction times for logistic growth with random catastrophes}
\author{Brandon H. Schlomann}
\email{bschloma@uoregon.edu}
\affiliation{Department of Physics and Institute of Molecular Biology, University of Oregon, Eugene, Oregon, 97405}
\date{\today}

\begin{abstract}
A central problem in population ecology is understanding the consequences of stochastic fluctuations.  Analytically tractable models with Gaussian driving noise have led to important, general insights, but they fail to capture rare, catastrophic events, which are increasingly observed at scales ranging from global fisheries to intestinal microbiota.  Due to mathematical challenges, growth processes with random catastrophes are less well characterized and it remains unclear how their consequences differ from those of Gaussian processes.  In the face of a changing climate and predicted increases in ecological catastrophes, as well as increased interest in harnessing microbes for therapeutics, these processes have never been more relevant.  To better understand them, I revisit here a differential equation model of logistic growth coupled to density-independent catastrophes that arrive as a Poisson process, and derive new analytic results that reveal its statistical structure.  First, I derive exact expressions for the model's stationary moments, revealing a single effective catastrophe parameter that largely controls low order statistics.  Then, I use weak convergence theorems to construct its Gaussian analog in a limit of frequent, small catastrophes, keeping the stationary population mean constant for normalization.  Numerically computing statistics along this limit shows how they transform as the dynamics shifts from catastrophes to diffusions, enabling quantitative comparisons.  For example, the mean time to extinction increases monotonically by orders of magnitude, demonstrating significantly higher extinction risk under catastrophes than under diffusions. Together, these results provide insight into a wide range of stochastic dynamical systems important for ecology and conservation.

\end{abstract}

\maketitle
\setlength\parskip{0pt}
\setlength\parindent{12pt}




\section{Introduction}
Stochastic fluctuations are important drivers of ecological and evolutionary processes \cite{levins1968evolution,landeBook,melbingerSREP_2015,fisher2014transition}.  Understanding their consequences is essential for ecological management, as well as for explaining observed patterns of biodiversity \cite{landeBook}.  Given that data is often limited, general principles of stochastic population dynamics derived from the mathematical analysis of minimal models can be immensely useful  \cite{landeBook,hastings2016timescales}.  For example, in classic work  \cite{beddington1977harvesting} Beddington and May derive for a stochastic logistic growth model how harvesting yields become less predictable as harvesting rates increase, a phenomenon that was suggested by historical fisheries data at the time \cite{schaefer1957study}.  Extensions of this analysis have led to threshold harvesting strategies that are proven optimal for a wide class of stochastic growth models that include extinction \cite{lande1995optimal}.  Beyond harvesting theory, analytically tractable models have led to diverse ecological and evolutionary insights \cite{levins1968evolution,melbingerSREP_2015,dickens2016analytically}.

In these types of analyses, stochasticity is often modeled by coupling growth to a Gaussian noise process, leading to stochastic differential equations that are amenable to well established tools from diffusion theory \cite{landeBook,karlin1981second}.  However, large, abrupt catastrophes are not captured by Gaussian models and are better modeled by discontinuous stochastic processes.  These catastrophes are increasingly observed in a variety of ecological systems.  On global scales, ecological catastrophes have already been observed as the result of rapid warming and are expected to become more frequent as the climate continues to change \cite{pershing2015slow}.  At the opposite extreme, the intestinal microbiomes of humans and other animals are observed to undergo abrupt compositional changes following perturbations, such as antibiotic treatments \cite{trosvik2015biotic, caporaso2011moving, relmanABX_2011,wilesPLOS2016}.  At all scales, efforts to understand and manipulate ecological systems would greatly benefit from general, quantitative principles of how perturbations and catastrophes shape population statistics.

\begin{figure*}
\centerline{
	\includegraphics[width =6.5 in]{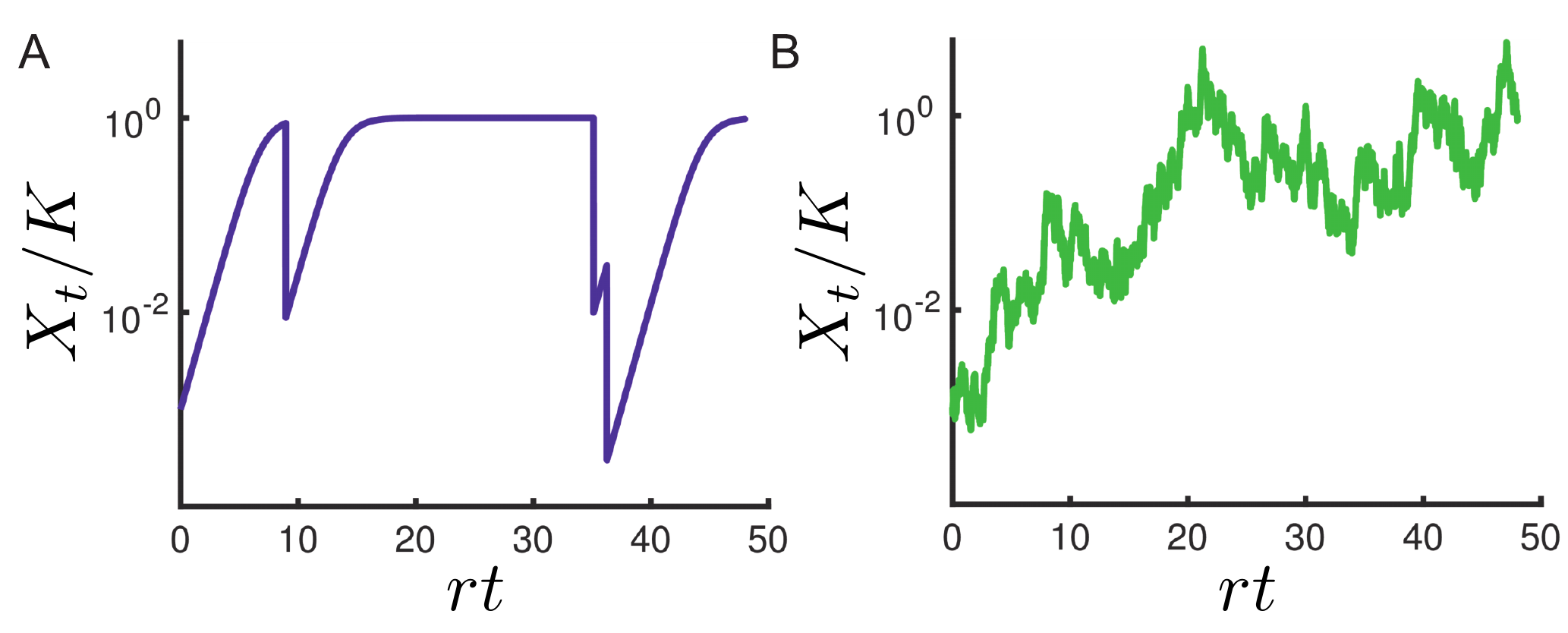}}
	\caption{{\bf{Sample paths of LRC and LES models.}}  {\bf{A:}}  A sample path from the LRC model.  Simulation parameters:  $r = 1$, $K = 10^4$, $\lambda = .07$, $f = 10^{-2}$, $dt = .01$.   {\bf{B:}}   A sample path from the LES model.  Simulation parameters:  $r = 1$, $K = 10^4$, $\sigma = .53$, $dt = .01$.      }
\end{figure*}

I address this issue here by  analytically and numerically studying a single-species model of logistic growth coupled to discontinuous, multiplicative jumps that arrive as a Poisson process, introduced in \cite{hanson1981} and referred to here as the Logistic Random Catastrophe (LRC) model (Figure 1A). Using the method of moment equations \cite{bover1978moment}, I derive exact expressions for the stationary moments of the population distribution, neglecting the possibility of extinction.  These results provide a direct look into the statistical structure of the LRC model, revealing a single, effective catastrophe parameter that largely controls ensemble statistics.  This effective parameter was recently observed empirically in computer simulations and aided the analysis of experimental data, but there was no theoretical basis for its existence \cite{wilesPLOS2016}.  

With this insight, I then turn to an old and fundamental problem: which dynamics, intermittent random catastrophes or continuous stochasticity, poses a higher risk of extinction?  For models of exponential growth up to a hard wall carrying capacity in the presence of either multiplicative Gaussian noise, called environmental stochasticity, or random, multiplicative Poisson catastrophes, Lande \cite{lande1993risks}  derives how the mean time to extinction scales as a power law in the carrying capacity for positive long-run growth rate, with the exponent depending on the details of the particular model.  This similarity in scaling behavior implies similar extinction risk in a qualitative sense, but it remains unclear how to construct a meaningful quantitative comparison, since the noise parameters of the two models describe distinct processes.

To circumvent this issue, I propose a method that treats the models not as distinct processes, but as extreme versions of the same process.  Using functional generalizations of the Central Limit Theorems \cite{jacod2013limit} and drawing inspiration from renormalization methods in theoretical physics \cite{peskin1995quantum,jona1975renormalization}, I analytically construct the diffusion analog of the LRC model, referred to here as the Logistic Environmental Stochasticity model (Figure 1B), in the limit of infinitely frequent, infinitesimal catastrophes, such that the stationary mean of the process remains constant.  In this way, the problem of quantitatively comparing two distinct models is traded for the more straightforward problem of computing statistics of one model as a function of parameters, specifically, along a particular limit in parameter space.  I apply this method to the comparison of extinction times and find that the mean time to extinction increases monotonically along this limit by orders of magnitude in a wide region of parameter space, implying significantly higher risk of extinction under random catastrophes dynamics in general.

Taken together, these results highlight the power of analytically tractable models of stochastic population dynamics.  The expressions derived here aid the analysis of experimental and observational data, inform the design of computer simulations, and reveal deep connections between distinct stochastic processes relevant for a wide range of ecological systems.

\begin{figure*}
\centerline{
	\includegraphics[width =6.5 in]{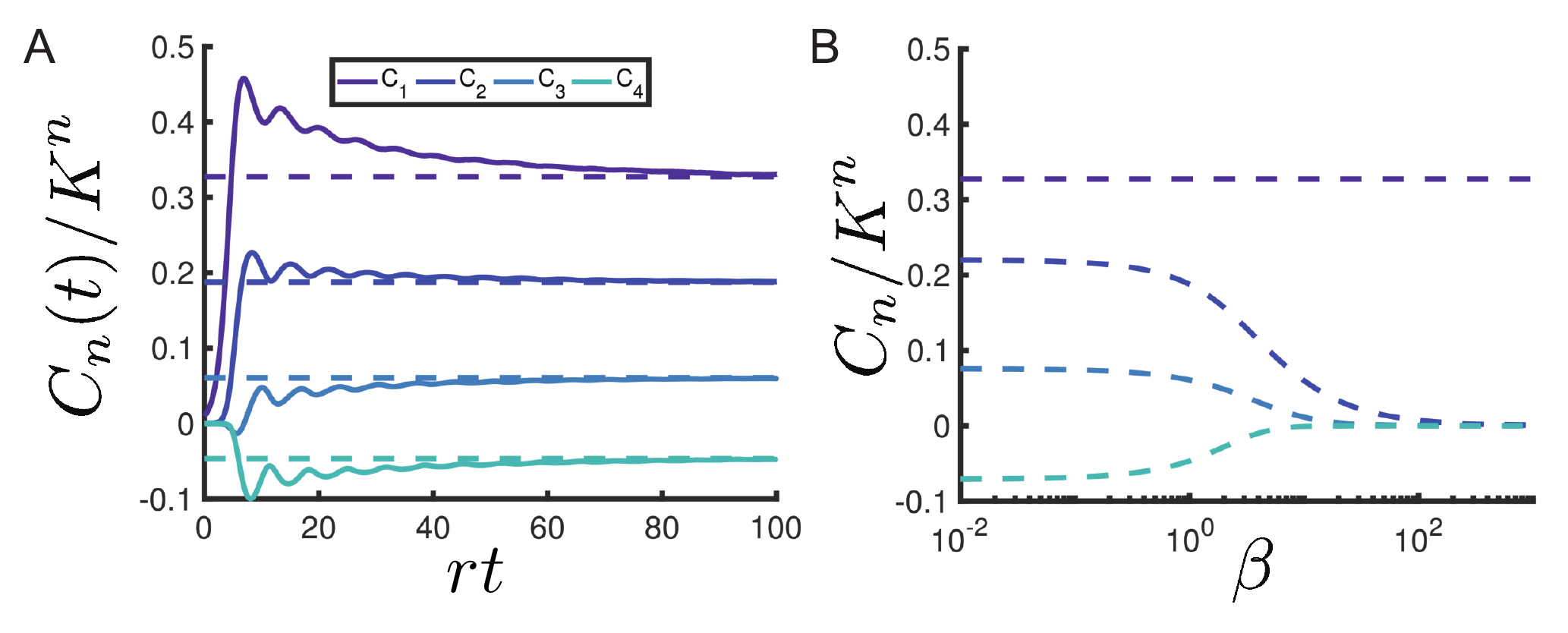}}
	\caption{\textbf{Analytic results reveal statistical structure of LRC model.  }\textbf{A:  }Analytic results for stationary cumulants agree with numerical simulations.  Time evolution of the first 4 cumulants, $C_n$, of the LRC model, computed numerically (solid lines).  Dashed lines indicate the asymptotic values predicted by the analytic results, with the cumulants computed from the moments given by equation (2).  Parameters:  $r = 1$, $K = 10^4$, $\lambda = .1$, $f = .0012$, $dt = .01$, $N_{trials} = 5\cdot 10^5$.  \textbf{B:  }Range of validity of $\lambda\ln f$ as an effective catastrophe parameter.  Parameters were scaled according to $\lambda' = \beta\lambda$ and $\ln f' = \beta^{-1}\ln f$ by dimensionless scale factor $\beta$.  Dashed lines are analytic results for first 4 stationary cumulants as a function of $\beta$.  Parameters same as in A.}
\end{figure*}

\section{Background on the Logistic Random Catastrophe model}

Hanson and Tuckwell \cite{hanson1981} introduce an ideal minimal model for the study of random catastrophes in isolation from additional complications:  Single-species logistic growth coupled to constant fraction catastrophes that arrive as a Poisson process, referred to here as the Logistic Random Catastrophe (LRC) model.  The LRC model can be written analytically as an It\^{o} Stochastic Differential Equation (SDE):
\be
dX_t = r X_t\left(1-\frac{X_t}{K}\right)dt  - (1-f)X_{t^-}dN_t.
\ee
\noindent The first term on the right hand side, of order $dt$, encodes deterministic logistic growth with growth rate $r$ and carrying capacity $K$.  The second term encodes random catastrophes with the use of a differential Poisson process, $dN_t$, which is equal to one if a catastrophe happens at time $t$ and zero otherwise.  Poisson catastrophes arrive with a constant probability per unit time, $\lambda$, and have a size set by $f$, the fraction of the population remaining after catastrophe.  The notation $X  _{t^-}dN_t$ indicates the It\^{o} integration convention \cite{hansonBook}.  By including logistic growth, the LRC model captures realistic density-dependent regulation; by including catastrophes of constant fraction, it captures the realistic feature that larger populations can experience larger losses, assuming that all individuals are equally susceptible to the disturbance.  Despite its simplicity, much about the statistical structure of the LRC model remains mysterious, due to the combined complications of the discontinuous Poisson process and nonlinear logistic growth.

\section{Results}
\subsection{Deriving exact expressions for LRC stationary moments}

I present here exact results for the stationary moments of the LRC model in absence of extinction, derived with the method of moment equations.  The method of moment equations turns a stochastic differential equation into an deterministic differential equation for the moment in question by averaging.  For nonlinear SDEs this results in a hierarchy of moment equations, in which each moment is coupled to higher moments, that generally cannot be solved exactly.  However, in the absence of extinction, this hierarchy reduces in the steady state to an algebraic recursion relation, which in the case of the LRC model is a simple relation between $\expec[X^{n+1}]$ and $\expec[X^{n}]$ (Appendix A).  This recursion relation can be iterated to express each moment just as a function of the mean.  Computing the mean independently (Appendix A) therefore determines all stationary moments:  


\be
\expec[X^n] = K^{n}\left(1 + \frac{\lambda}{r}\ln f\right)\prod^{n-1}_{m=1}\left(1 - \frac{\lambda(1-f^m)}{mr}\right),
\ee

\noindent from which expressions for the stationary mean and variance are readily obtained,
\be
\expec[ X]_{LRC} = K\left(1+\frac{\lambda}{r}\ln f\right),
\ee
\be
 \Var[X]_{LRC} = K^2\frac{\lambda}{r}\left(-\ln f-(1-f)\right)\left(1+\frac{\lambda}{r}\ln f\right)
\ee
\noindent (recall that $f \in (0,1)$, so $\ln f$ is negative for $ f < 1$).

These results agree well with simulations, as shown in Figure 2A in the form of cumulants \cite{broca2004cumulant}, which generally provide more intuitive information than moments.  The solid lines show the time evolution of the first 4 cumulants, $C_n$, of the LRC model, computed via stochastic simulation of the Poisson process with no absorbing state representing extinction (Materials and Methods).  The dashed lines are the analytic results, computed from the expressions for the moments in equation (2)  \cite{broca2004cumulant}.  Each cumulant asymptotes to the analytic value.

\subsection{A single, effective catastrophe parameter largely controls LRC moments}

These analytic results suggest that the parameter combination $\lambda\ln f$ plays an important role in determining population statistics.  To investigate its role, I computed the response of the first four stationary cumulants of the LRC model to simultaneous, reciprocal scaling of $\lambda$ and $\ln f$ via a dimensionless scale factor, $\beta$.  In regions of parameter space where statistics depend only on the effective parameter $\lambda\ln f$, curves of cumulants as a function of $\beta$ will be flat.  The results are shown Figure 2B for $\beta$ ranging from $10^{-2}-10^3$, with catastrophe parameters scaled as $\lambda' = \beta\lambda$ and $\ln f' = \beta^{-1}\ln f$.  Stationary moments were computed for each value of $\beta$ using the analytic results derived above and converted to cumulants \cite{broca2004cumulant}.  The stationary mean is invariant under this scaling, as indicated by equation (3).  Higher order cumulants are approximately invariant for low values of $\beta$, which correspond to rare, large catastrophes, but decay to zero for large values of $\beta$, which correspond to frequent, small catastrophes.  

The large $\beta$ limit is a dynamic analog to the law of large numbers, in which the Poisson process that drives the LRC model tends to its average value.  Recall that for a Poisson process, all cumulants are equal to the mean, $\lambda t$, just as all cumulants of a Poisson distribution are equal to the mean.  The $n^{th}$ cumulant of the scaled process $(1-f)N_t$ is therefore $(1-f)^n\lambda t$. The limit $\beta \to \infty$ corresponds to $\lambda'\to\infty$ and $f'\to 1$, such that $\lambda'\ln f'$ is constant.  In this limit, $-(1-f)$ is well approximated by $\ln f$, so, higher cumulants of the Poisson process decay as $\beta^{-(n-1)}$, resulting in a deterministic model.   

The effective parameter $\lambda\ln f$ has an intuitive interpretation:  It is the correction to the long-run growth rate due to catastrophes, and has been previously identified as an important quantity in a variety of related models \cite{bao2011competitive,hansonBook,lande1993risks}.  Its existence has important consequences for analyzing experimental data.  As was done in \cite{wilesPLOS2016}, fitting ensemble statistics with the effective catastrophe parameter reduces the number of parameters that needs to be estimated from data.  In fact, attempting instead to fit both the rate $(\lambda)$ and size $(f)$ independently results in highly unconstrained parameter estimates  \cite{wilesPLOS2016} and should be avoided.  The analytic results derived here put the use of the effective parameter, $\lambda\ln f$, on firmer ground and explicitly delineate the range of its validity.  

 \begin{figure*}
\centerline{
	\includegraphics[width =6.5in]{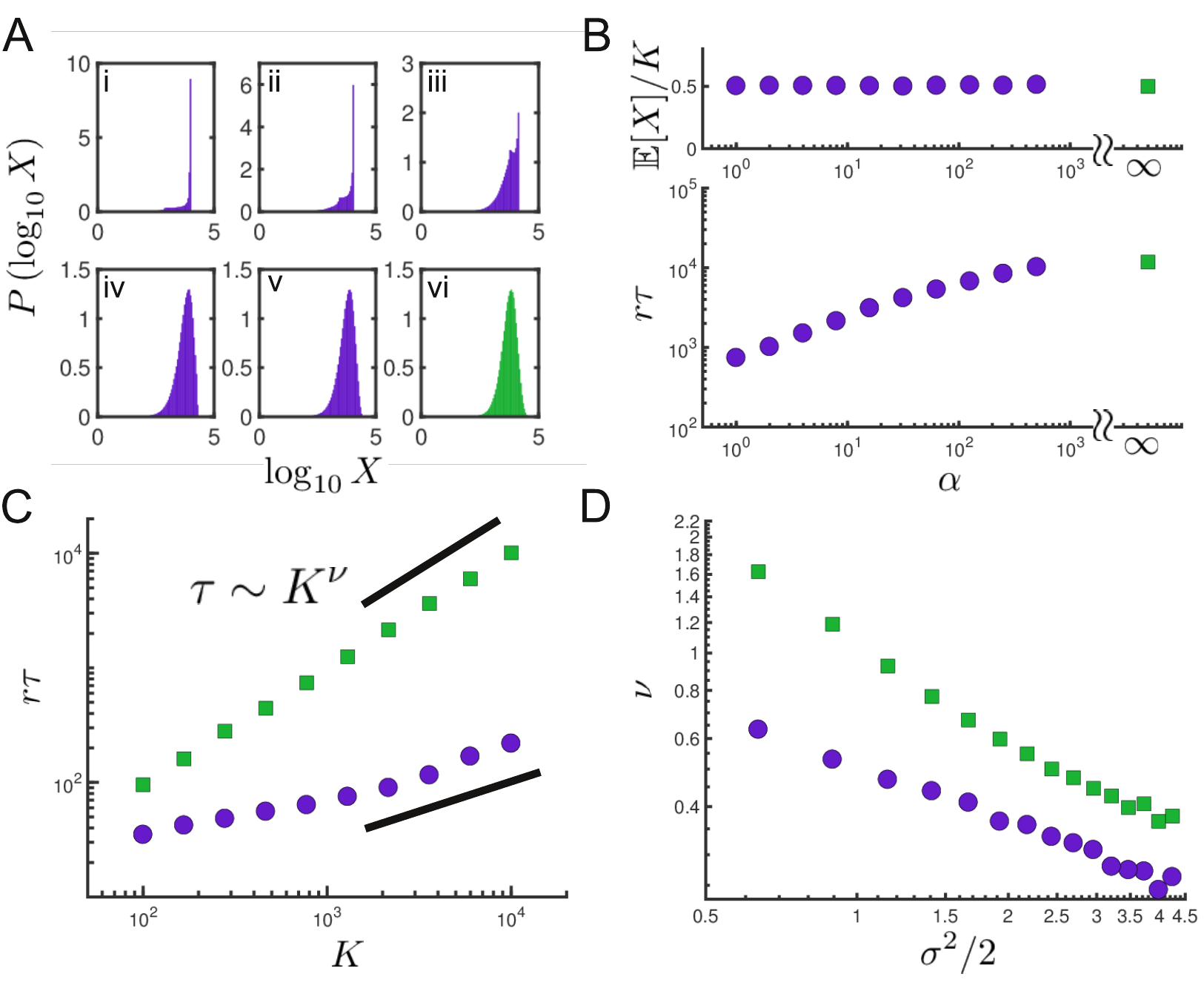}}
		\caption{\textbf{The diffusion limit and extinction times.  }\textbf{A:  }Smoothly transforming the stationary distribution of LRC model to that of LES model in the diffusion limit.  LRC model (no extinction) was simulated for $T_{max} = 300$ units of inverse growth rate for 5 values of the scale parameter $\alpha$, rescaling parameters according to equation (7).  Frames i-v depict the stationary distribution of $\log X$ (for visual clarity) for $\alpha = 1, 2.94, 8.66, 25.49, 75.00$ respectively, estimated from $10^6$ paths.  Frame vi depicts the stationary distribution of $\log X$ for the target LES model.  Parameters:  $r = .68$, $K = 6800$, $\lambda = .1$, $\sigma = .8$, $\ln f = -\sigma/\sqrt{\lambda}$, $dt = .01$.  \textbf{B: }  Mean time to extinction increases as the LRC model is morphed into the LES model (bottom), despite the stationary mean remaining constant throughout the transformation (top).  Parameters:  Same as in A but $\alpha$ ranges logarithmically from 1 to 500 and $\sigma = 1$.  \textbf{C:  } Mean times to extinction for the beginning (purple circles) and end (green squares) points of the diffusion limit as a function of carrying capacity.  Parameters:  $r = 1$, $\lambda = .1$, $f = .01$, $N_{trials} = 5000$, $x_0 = 10$, $x^* = 1$.  For LES endpoints, $\sigma^2 = \lambda \ln^2f$.  \textbf{D:  }. The exponent of this power law, obtained by linear regression, as a function of effective noise strengths.  Parameters:  $r = 1$, $f = .01$, $N_{trials} = 5000$, $\lambda$ is varied from $.06$ to $.4$.  For each value of $\lambda$, $\tau$ vs $K$ is computed for 10 values of $K$ ranging from 100 to 10000.  The last 5 values are used to compute $\nu$.  For the LRC model, the $x$-axis corresponds to $\sigma^2 = \lambda\ln^2f$.  The exponent appears to asymptotically follow a power law in $\sigma^2$, consistent with \cite{lande1993risks}}
\end{figure*}

\subsection{The diffusion limit shows that random catastrophes pose higher extinction risk than environmental stochasticity}

I now consider the problem of quantitatively comparing extinction risks in the LRC model and its environmental stochasticity analog, referred to here as the Logistic Environmental Stochasticity (LES) model.  The LES model can be written as an SDE,
\be
dX_t = rX_t\left(1-\frac{X_t}{K}\right) + \sigma X_{t^-} dB_t,
\ee
\noindent with $B_t$ the standard Brownian motion process \cite{hansonBook,karlin1981second}, whose intervals are independent, Gaussian distributed variables with $\expec[B_t] = 0$ and $\Var[B_t] = t$,  and $\sigma$ setting the strength of the noise.  Historically, there has been no obvious way of quantitatively comparing extinction risk between the two models across parameter space, since the noise parameter $\sigma$ and the catastrophe parameters, a rate $\lambda$ and size $f$, describe distinct, model-specific processes \cite{lande1993risks, landeBook, shaffer1987minimum}.

To circumvent this issue, I propose an approach in which the LES model is viewed not as a distinct process, but as a special case of the LRC model.  This notion has been expressed qualitatively for decades \cite{shaffer1987minimum,lande1993risks}, but, to my knowledge, has never been made explicit.  This can be done using a functional generalization of the Central Limit Theorem (CLT) \cite{jacod2013limit}, which says that fluctuations of the Poisson process about its mean converge in distribution to Brownian motion in the limit of infinite jump rate and infinitesimal jump size.  In Appendix B, it is shown that the relevant limits are 
\be
\ln f(N_t - \lambda t) \xrightarrow{\lambda \to \infty, \text{ } f \to 1} \sqrt{\lambda} \ln fB_t,
\ee
\noindent such that $\sqrt{\lambda}\ln f$ is constant.  Consequently, the mean drift of the scaled Poisson process diverges as $\sqrt{\lambda}$, the variance remains finite and all higher cumulants go to zero.  These are functional analogs to what happens when a Poisson distribution limits to a Gaussian in the classical CLT.   In this case, the diverging drift - which is proportional to effective catastrophe parameter $\lambda\ln f$ discussed above - is a manifestation of the fact that catastrophes are unidirectional, whereas noise in the LES model is bidirectional.  To obtain a non-trivial limiting process, this drift must be subtracted off manually before taking limits.  This subtraction can be absorbed into a rescaling of the growth rate and carrying capacity, similar to renormalization methods in theoretical physics \cite{peskin1995quantum}, such that the final transformation from the LRC model to the LES model involves rescaling all four LRC model parameters.

The complete transformation from the LRC to LES model will be parameterized by a dimensionless scale parameter, $\alpha$.  The prescription is as follows.  Start from an LRC model together with a target LES noise strength $\sigma$ and fix $\lambda\ln^2 f = \sigma^2$.  Then transform the LRC parameters according to
\begin{equation*}
\lambda' = \alpha \lambda,\text{\hspace{.4cm}}\ln f' = -\sigma/\sqrt{\lambda'},
\end{equation*}
\begin{equation}
r' = r\left(1- \frac{\lambda'\ln f'}{r}\right),\text{\hspace{.2cm}} K' = K\left(1  -\frac{\lambda'\ln f'}{r}\right).
\end{equation}
\noindent It is shown analytically in Appendix B that in the limit $\alpha \to \infty$, the LRC model $LRC(r',K',\lambda',f')$ gets mapped to an LES model $LES(r_{ES},K_{ES},\sigma)$, with $\sigma = \lambda\ln^2f$, $r_{ES} = r(1+(2r)^{-1}\sigma^2)$, and $K_{ES} = K(1+(2r)^{-1}\sigma^2)$.  In this way, the stationary mean of the process without extinction remains constant throughout this transformation, fixed at $K$.  This is chosen as a convenient way to normalize the effects of noise.  By adding constant offsets, the transformation can be tuned to preserve other properties (Appendix B).  This transformation is shown visually in Figure 3A, which depicts numerical results for the stationary distribution of the LRC model (in log variables for visual reasons) being transformed with $\alpha$ increasing on the interval $(1,75)$ (Materials and Methods).  The distribution approaches that of the target LES model, shown in green in panel (vi) .  

With this transformation, the question of relative risks of extinction under the LRC model and the LES model was revisited.  The mean time to extinction, $\tau$, was computed via stochastic simulation of the LRC model for various values of $\alpha$ (Figure 3B, bottom), rescaling LRC model parameters according to equation (7) for each $\alpha$ (Materials and Methods).  The LRC model extinction time (purple circles)  increases with increasing $\alpha$ and asymptotes to the LES model extinction time (green square).  Computed numerically, the stationary population mean in the absence of extinction does indeed remain constant throughout the transformation (Figure 3B, top).  The conclusion is again that there exists a significantly higher risk of extinction under random catastrophe dynamics than under environmental stochasticity dynamics.  

This conclusion is robust across parameter space.  Plotting the beginning and end points of the curve in Figure 3B for various values of carrying capacity reproduces the asymptotic power law behavior described by Lande for simpler models \cite{lande1993risks} (Figure 3C), though the exponents obtained by linear fitting (Materials and Methods) are smaller for the LRC model across a wide range of effective noise strengths (Figure 3D).  Note that because the growth rate is rescaled in this procedure, as long as the original growth rate $r$ is positive, the long-run growth rate \cite{lande1993risks} is positive for all values of $\sigma$.  The conclusion is also insensitive to the initial starting population, $x_0$, as the mean time to extinction becomes independent of $x_0$ above a critical threshold (see \cite{landeBook,hanson1981} and Supplementary Figure 1).  In addition to this method based on the diffusion limit, an alternative approach, in which the stationary means of the LRC and LES models equated simply by mapping $\sigma^2 = -2\lambda\ln f$, leads to the same conclusion (Supplementary Figure 2, Appendix C).

\section{Discussion}

This work presented new results for the Logistic Random Catastrophe (LRC) model, a model that serves both as a foundation for understanding the ecological consequences of random catastrophes and as an empirical model that describes real data \cite{hanson1981,wilesPLOS2016}.  Exact analytic results for its stationary moments were derived using the method of moment equations.  These expressions revealed that ensemble statistics are largely controlled by a single parameter that combines the average catastrophe rate and size, which is both a fundamental insight into the model's statistical structure and a useful result for the analysis of ecological data \cite{wilesPLOS2016}.  They also revealed the similarity in structure between the LRC model and its Gaussian noise counterpart, the Logistic Environmental Stochasticity (LES) model, which was exploited to construct the latter as a limit of the former.  The mean time to extinction increased monotonically along this limit by orders of magnitude in relevant regions of parameter space, indicating higher extinction risk under random catastrophe dynamics in general.  This has implications for the prioritization of conservations efforts in the face of different types of stochasticity.  In addition, given that large fluctuations appear to be intrinsic to intestinal microbiota \cite{trosvik2015biotic, caporaso2011moving,wilesPLOS2016}, the enhanced extinction risk of reported here may be important for understanding the evolution of functional redundancy across symbiotic taxa and of host biochemical networks that sense fluctuating microbial products.

The relationship between the LES model and the LRC model constructed here has multiple interpretations, which together provide useful intuition.  For one, the LES can be thought of as existing in a subset of the LRC parameter space, namely, as a particular limit in the direction $\lambda\to\infty$, $r\to\infty$, $K\to\infty$, $f\to 1$.  This is analogous to how exponential growth can be considered a special case of logistic growth with infinite carrying capacity.  Alternatively, if the Central Limit Theorem is interpreted as dictating the fixed point of a coarse graining procedure, in which case it becomes an example Renormalization Group methods from statistical physics (albeit a simple one) \cite{jona1975renormalization}, its application to stochastic processes can be interpreted as a statement about iterative temporal coarse graining: any stochastic process with independent increments, zero mean, and finite variance resembles Brownian motion when viewed on long time scales with appropriate rescaling.  As a result, the LRC model flows to the LES model.  

Computing statistics along this limit lead to quantitative insight into extinction risks.  This method is readily applied to the study of other statistics of the LRC model.  It can also be easily adapted to other Markov models, including multi-species models \cite{bao2011competitive,jacod2013limit}.  The conditions for convergence are given in \cite{jacod2013limit} and amount to reasonable boundedness conditions on the transition kernel of the Markov process.  These generalizations allow for more computations, analogous to extinction times in Figure 3B, that could provide useful insight.  For one example, it would be useful to revisit optimal control problems relevant for ecological management in the presence of random catastrophes, such as the harvesting strategies for fisheries considered in \cite{hanson1998optimal}, and study how optimal policies evolve when discontinuous jumps limit to continuous environmental stochasticity.  For another, evolutionary studies of bet hedging in the presence of catastrophes \cite{visco2010switching}  could be directly mapped to the analogous problem in the presence of continuous noise \cite{melbingerSREP_2015}, connecting ecological and evolutionary dynamics relevant for a wide variety of systems.

\section{Materials and Methods}
 \noindent All code was written in MATLAB and is available at  \url{https://github.com/bschloma/lrc}.
 
\subsection{LRC and LES model simulations}Sample paths of the differential Poisson process were generated as Bernoulli trials \cite{hansonBook}.  These paths were then used in the numerical integration of the LRC model.  For all calculations except for the diffusion limit calculations in Figure 3, the logistic growth equation was integrated with the Euler method between jump times, at which the population was reduced by a factor of $f$.  The LES model was integrated with a straightforward application of the Milstein method \cite{mil1975approximate}.

\subsection{The diffusion limit}
In the diffusion limit, jump sizes approach the size of deterministic growth in one numerical timestep.  So, the deterministic contribution of order $\Delta t$ must be retained, resulting in a more straightforward Euler-type integration scheme.  In this case, an adaptive timestep is used, scaling $\Delta t' = \Delta t/\sqrt{\alpha}$, identically to $\ln f$, which sets the size of the jump.  This scaling will lead to numerical artifacts when the probability of catastrophe in one timestep, $\lambda'\Delta t'$, approaches unity.   Since $\lambda' = \alpha \lambda$, this will occur at $\alpha_c \sim (\lambda\Delta t)^{-2}$, and so can be put off by starting with a sufficiently small time step.  

\subsection{Extinction times}
Extinction times were computed by straightforward stochastic simulation, following population trajectories from an initial population, $x_0$ until they reached the extinction threshold, $x^*$.  To extract the exponent, $\nu$, of the asymptotic relationship $\tau \sim K^{\nu}$, a linear fit to log-transformed variables was done for the larger half of the carrying capacity values, typically 5 data points.

\section{Acknowledgements}
I thank Raghuveer Parthasarthy, Pankaj Mehta, and David Levin for helpful feedback.  Research reported in this publication was supported by the NIH as follows: by the NIGMS under award number P50GM098911 and by training grant T32 GM007759.

\onecolumngrid
\appendix
\section{Detailed calculation of stationary moments}
In this section an expression for the $n^{\text{th}}$ stationary moment for the LRC model is derived.  The approach is analogous for the LES model and since the results are already known \cite{engen2000}, a detailed derivation isn't given, though one remark is made on the application of this method to diffusion processes.

\subsection{LRC model}

Before beginning, the chain rule for jump processes \cite{hansonBook} is stated without proof, for reference.  Let $X_t$ by a general process given by
\be
dX_t = f(X_t,t)dt + h(X_{t^-},t^-)dN_t
\ee

\noindent with $f$ and $h$ deterministic functions, $N_t$ a Poisson process with rate $\lambda$, and $t^-$ denoting the It\^{o} convention as in the main text.  Further let $Y_t \equiv F(X_t,t)$ be a transformed process.  Then $Y_t$ is governed by
\be
dY_t = \left(\partial_tF(X_t,t) + f(X_t,t)\partial_{X_t}F(X_t,t)\right)dt\\ + \Delta Y_{t^-}^{jump}dN_t
\ee

\noindent with $\Delta Y_{t^-}^{jump} \equiv F(X_{t^-} + h(X_{t^-},t^-)) - F(X_{t^-},t^-)$.

Now recall the LRC model,
\be
dX_t = r X_t\left(1-\frac{X_t}{K}\right)dt  -(1-f)X_tdN_t.
\ee

\noindent The first step is to change variables to $X^n_t$ using the stochastic chain rule for jump SDEs.  The result is
\be
dX^n_t = nr X^n_t\left(1 -\frac{X_t}{K}\right)dt -(1-f^n)X^n_tdN_t.
\ee

\noindent Then, each term in this SDE is averaged.  The expectation of $X^n_{t^-}dN_t$ can be factored:  $\expec[X^n_{t^-}dN_t] = \expec[X^n_{t^-}]\expec[dN_t] = \expec[X^n_{t^-}]\lambda dt$.  Intuitively, this is because the two processes appear mutually independent.  The Poisson process has independent increments, and since the It\^{o} convention was used, $X^n_{t^-}$ is independent of $N_t$, which occurs in the future.  This is certainly true for a discrete time model, but care must be taken in the continuous limit.  

A more rigorous argument can be made using the Dominated Convergence Theorem.  The case $n=1$ is considered without loss of generality.  Consider $X_j$, a discrete partition of the continuous time process $X_t$, such that $X_j \to X_t$ in probability.  Then, sums of $X_j$ converge in probability to integrals, in particular,
\be
\sum_j X_{j-1} \Delta N_j \to \int_T X_{t^-} dN_t,
\ee

\noindent where $\Delta N_j$ is a partition of the Poisson process.  The Dominated Convergence Theorem says that if $X_t$ is dominated by an integrable function on the interval $T$,
\be
\expec\left [\sum_j X_{j-1} \Delta N_m\right] \to \expec\left[\int_T X_{t^-} dN_t\right]
\ee

\noindent in probability.  Since populations in the LRC model are bounded by the carrying capacity for all time, this is always valid.  The expectation of the sum is straightforward, leading to the result,
\be
\expec\left[\int_T X_{t^-} dN_t\right] = \int_T\expec[X_{t^-}]\lambda dt,
\ee

\noindent from which the infinitesimal version follows as a special case.

Factoring the expectation results in an ODE for the $n^{th}$ moment.  In the steady state, this becomes the recursion relation
\be
\expec[X^{n+1}] = K\left(1 - \frac{\lambda(1-f^n)}{nr}\right)\expec[X^{n}].
\ee

\noindent Defining 
\be
c_n \equiv \left(1 - \frac{\lambda(1-f^n)}{nr}\right),
\ee

\noindent the $n^{th}$ moment can be expressed in terms of the mean as
\be
\expec[X^n] = K^{n-1}\left(\prod^{n-1}_{m=1}c_m \right)\expec[X].
\ee

To complete the recursion relation, the mean must be computed independently.  This is accomplished by changing variables to $\ln X_t$ using the chain rule for jump processes:
\be
d\ln X_t = r\left(1 -\frac{X_t}{K}\right)dt +\ln f dN_t,
\ee

\noindent which in the steady state gives an expression for the stationary mean,
\be
\expec[X] = K\left(1 + \frac{\lambda}{r}\ln f\right).
\ee

\noindent Plugging this back into equation (A10) gives the final result
\be
\expec[X^n] = K^{n}\left(1 + \frac{\lambda}{r}\ln f\right)\prod^{n-1}_{m=1}\left(1 - \frac{\lambda(1-f^m)}{mr}\right).
\ee

\noindent Evaluating this equation for $n=2$ leads to the expression for the variance in the main text:
\be
 \Var[X]_{LRC} = K^2\frac{\lambda}{r}\left(-\ln f-(1-f)\right)\left(1+\frac{\lambda}{r}\ln f\right).
\ee

\subsection{LES model}
The derivation is analogous for the LES model, except that It\^{o}'s chain rule for diffusion processes is used.  Since the results are already known \cite{engen2000}, derived with traditional methods, a detailed computation will not be given.  However, one remark worth making concerns the expectation of $X_{t^-}dB_t$.  The intuitive argument outlined for the LRC model - that since the It\^{o} convention was employed the expectation of the product can be factored - gives the correct answer in this case, but is in fact not generally valid.  Essentially, for processes governed by equations of the form 
\be
dX_t = f(X_t,t)dt + g(X_{t^-},t^-)dB_t,
\ee

\noindent the integral $\int g(X_{t^-},t^-)dB_t$ can acquire non-zero expectation if the function $g$ grows too quickly.  A classic example is the CEV model of quantitative finance \cite{linetsky2010constant}, which is of the form $f(X_t,t) = X_t$ and $g(X_t,t) = X^{\gamma}_t$ for $\gamma > 1$.  However, one can use the fact that the exponential version of the LES model, i.e. $K \to \infty$, is a well known SDE for which $\expec\left[\int X_{t^-}dB_t\right]=0$.  This model is known as Geometric Brownian Motion and describes asset prices in the Black-Scholes model of quantitative finance \cite{linetsky2010constant}.  Since paths of the exponential model almost surely dominate paths of the LES model, $\int X_{t^-}dB_t$ for the LES model inherits the martingale property from the exponential case, which implies zero expectation.

Following the same procedure as for the LRC model, factoring expectations of $X^n_{t^-}dB_t$, results in

\be
\expec[X^n]_{LES} = K^{n}\left(1-\frac{\sigma^2}{2r} \right)\prod^{n-1}_{m=1}\left(1+\frac{(m-1)\sigma^2}{2r} \right).
\ee

\noindent Special cases of this include

\be
\expec[X]_{LES} = K\left(1-\frac{\sigma^2}{2r}\right)
\ee

\noindent and

\be
\Var[X]_{LES} = \frac{K^2\sigma^{2}}{2r}\left(1-\frac{\sigma^2}{2r}\right).
\ee

\section{The diffusion limit and the Central Limit Theorem}

This section contains details of the construction of the LES model from the LRC model in the limit of infinitely frequent, infinitesimal catastrophes, referred to here as the diffusion limit.  The complete transformation involves all four LRC model parameters and is specified as follows.  Let $\alpha$ be a scale parameter, $LRC(r,K,\lambda,f)$ an LRC model, and $\sigma$ be the target noise-strength parameter of the limiting LES model.  Fix $\lambda\ln^2 f = \sigma^2$, and scale
\begin{equation*}
\lambda' = \alpha \lambda,\text{\hspace{.4cm}}\ln f' = -\sigma/\sqrt{\lambda'}
\end{equation*}
\begin{equation*}
r' = r\left(1- \frac{\lambda'\ln f'}{r}\right) = r\left(1+ \frac{\sigma\sqrt{\lambda}}{r}\sqrt{\alpha}\right),
\end{equation*}
\begin{equation}
K' = K\left(1  -\frac{\lambda'\ln f'}{r}\right) = K\left(1+ \frac{\sigma\sqrt{\lambda}}{r}\sqrt{\alpha}\right).
\end{equation}

\noindent The claim is that in taking the limit $\alpha \to \infty$, the LRC model $LRC(r',K',\lambda',f')$ gets mapped to an LES model $LES(r_{ES},K_{ES},\sigma)$, with $\sigma^2 = \lambda\ln^2f$, $r_{ES} = r(1+(2r)^{-1}\sigma^2)$, and $K_{ES} = K(1+(2r)^{-1}\sigma^2)$, such that the stationary means of both models are equal. I first motivate the form of this transformation, which involves all four LRC model parameters, by studying the behavior of the stationary moments.  I then show how the precise form of these limits, namely $\lambda\to\infty$, $f\to 1$, such that $\lambda\ln^2f\to$ const., follows from functional generalizations of the Central Limit Theorem (CLT), in which a scaled, compensated Poisson process limits to Brownian motion.  Finally, I show analytically how the full transformation maps the LRC model into the LES model.  

\subsection{Motivation}

As discussed in the main text, taking the limits $\lambda\to\infty$, $f\to 1$, such that $\lambda\ln f\to$ const. is analogous to the law of large numbers, leading to a deterministic limit.  The correct limits instead are $\lambda\to\infty$, $f\to 1$, such that $\lambda\ln^2f\to$ const., which I show below is analogous to the CLT.  To motivate the final four parameter transformation, let us first consider the behavior of the LRC variance under these limits:

\bea
\Var[X] &=& K^2\frac{\lambda}{r}(-\ln f -(1-f))\left(1+\frac{\lambda}{r}\ln f\right)\nonumber\\ 
&\xrightarrow{\text{limits}}&K^2\frac{\lambda \ln^2 f}{2r}\left(1+\frac{\lambda}{r}\ln f\right)\nonumber\\
&=&  K^2\frac{c^2}{2r} -K^2\frac{c^3}{r^2}\sqrt{\lambda}.
\eea
\noindent with $c = \text{ const } = -\sqrt{\lambda}\ln f$.  In taking the limit $f\to1$, the relation $(-\ln f -(1-f)) \to 2^{-1}\ln^2f$ was used, based on a $2^{\text{nd}}$ order Taylor expansion.  

The variance diverges, but a part of it remains finite.  The finite piece of the variance in this limit is exactly the variance of an LES model with $\sigma^2 = \lambda\ln^2 f$ and increased growth parameters $K_{ES} = K(1+(2r)^{-1}\sigma^2)$, and $r_{ES} = r(1+(2r)^{-1}\sigma^2)$.  Looking at the behavior of the mean in this limit leads to the same conclusion.  This suggests that this limit does take the LRC model into an LES model, but one that is accompanied by a noise-induced drift that diverges as $\sqrt{\lambda}$.  This is divergence should be expected, as it reflects the unidirectionality of jumps in the LRC model, which is absent in the LES, analogous to the divergence of the mean of a Poisson distribution when it limits to a Gaussian.  To obtain a non-trivial limiting process, this drift needs to be subtracted off, for example, by adding a term $-\lambda\ln f dt$ to the LRC model SDE.  This is equivalent to rescaling the growth rate and carrying capacity each by a factor of $(1-r^{-1}\lambda\ln f)$, leading to the full four parameter transformation.

\subsection{Functional Central Limit Theorems}

The form of the limits  $\lambda\to\infty$, $f\to 1$, such that $\lambda\ln^2f\to$ const., is a direct consequence of the CLT.  The classical CLT says that given a set of $n$ random variables, $\{\xi_j\}$, that are identically and independently distributed (i.i.d.) with mean $\mu$ and finite variance $\sigma^2$, the sum of the deviations of these variables from their mean, when rescaled by $\sqrt{n}$, tends in distribution to a Gaussian variable as $n\to\infty$:

\be
\lim_{n\to\infty}\frac{\sum_j\xi_j -n\mu}{\sqrt{n}} = \eta \sim \mathcal{N}(0,\sigma^2)
\ee

\noindent where $\mathcal{N}(\mu,\sigma^2)$ is a Gaussian distribution with mean $\mu$ and variance $\sigma^2$.  The condition that the variables $\xi_j$ follow identical distributions can be relaxed, but we focus on this restricted case here.  

A vast body of mathematical literature concerns the construction of generalization of the CLT to stochastic processes.  One important generalization, which we will employ in the study of the Poisson process, is Donsker's theorem \cite{jacod2013limit}.  Donsker's theorem dictates the limit of a sequence of stochastic process, $X^{(n)}_t$, constructed from sums of i.i.d. random variables $\tilde{\xi}_j$ with zero mean and finite variance via

\be
X^{(n)}_t \equiv \frac{1}{\sqrt{n} }\sum_{j=1}^{[ns]}\tilde{\xi}_j\text{, \hspace{.2cm}}ns \equiv t.
\ee

\noindent Here we have introduced time as multiples of a unit $s$, such that $t = ns$, and $[...]$ denotes the integer part.  Donsker's theorem says that as $n\to\infty$ with $s\to 0$ such that $ns\to t$ for arbitrary $t$, the processes $X^{(n)}_t$ converge in law to Brownian motion,
\be
X^{(n)}_t \to B_t.
\ee

\noindent Donsker's theorem can be used to show the convergence of the compensated Poisson process to Brownian motion in particular limits.  The idea is to write the Poisson process as a sum of intervals which themselves are i.i.d. random variables that meet the criteria for Donsker's theorem, and then scale the jump size and rate in the ways that map onto the $n\to\infty$ limit.  This approach is based on a method known as finite dimensional convergence, which is only applicable to processes with independent increments \cite{jacod2013limit}.  

Consider breaking a scaled, compensated Poisson process, $\epsilon \tilde{N}_t$ into a sum of finite intervals,

\be
\epsilon \tilde{N}_t = \epsilon\sum_{j=1}^{[ns]}\Delta \tilde{N}_j.
\ee

\noindent From inspection, we see that the appropriate mapping is $\lambda \to n\lambda$, $\epsilon = \sigma\lambda^{-1/2}$, in which case

\be
\epsilon \tilde{N}_t = \sigma\sum_{j=1}^{[ns]}\frac{\Delta \tilde{N}_j/\sqrt{\lambda}}{\sqrt{n}}.
\ee

\noindent  Donsker's theorem can then be applied with $\tilde{\xi}_j \equiv \Delta \tilde{N}_j/\sqrt{\lambda}$, resulting in

\be
\epsilon \tilde{N}_t  \xrightarrow{\lambda \to \infty, \text{ } \epsilon \to 0} \sqrt{\lambda}\epsilon B_t.
\ee

In the LRC model, collapse size is forced to zero by taking $f\to 1$.  This still leaves room for how exactly $\lambda$ and $f$ should map on to $\sigma$.  One choice would be to take $\epsilon = -(1-f)$, such that $\sigma^2 = \lambda (1-f)^2$.  In this case, a quick calculation shows that the limiting process would be an LES model with unchanged growth parameters, $(r,K)$, and consequently a reduced stationary mean of $K(1-(2r)^{-1}\sigma^2)$, whereas the original LRC process, after removing the divergence of $\lambda\ln f$, has a stationary mean of $K$.  Alternatively, one could take  $\epsilon = \ln f$, such that $\sigma^2 = \lambda \ln^2f$.  This is the case examined above, which results in an LES model with an unchanged stationary mean, but altered growth parameters.  Since our present goal is to normalize the effect of noise to construct a fair comparison of extinction risk, the latter choice is more appropriate.

\subsection{Convergence of LRC to LES}

I now discuss the convergence of the LRC model to the LES model via the convergence of the Poisson process to Brownian motion discussed above.  General conditions for the convergence of a pure jump Markov process to a diffusion are given in \cite{jacod2013limit}.  Rather than verify these general conditions here, I'll take a more intuitive approach that exploits the simplicity of the present models and uses the fact that both the LRC model and the LES model possess unique, strong solutions, as follows from special cases of a general result derived in \cite{bao2011competitive}.  This allows us to uniquely define a sequence of processes $X'_t(\alpha) \equiv F[\tilde{N}_t(\alpha); r(\alpha),K(\alpha),\lambda(\alpha),f(\alpha)]$, where $F$ is the solution to the LRC model depending on parameters $r$, $K$, $\lambda$, and $f$, and the limit of the sequence $X^*_t \equiv \lim_{\alpha\to\infty} X'_t(\alpha)$.   Existence and uniqueness of solutions to both models allows us in principle to take the limit and then invert the solution, recovering a diffusion SDE.  In practice, we can take the limit directly in the context of the LRC model SDE.  
To begin, recall the LRC model,
\be
dX_t = r X_t\left(1-\frac{X_t}{K}\right)dt -(1-f)X_{t^-}dN_t.
\ee

\noindent Let us expand $(1-f)$ in powers of $\ln f$ to second order and write the Poisson process in terms of its mean and compensated process.

\be
dX_t = r X_t\left(1-\frac{X_t}{K}\right)dt +\left(\ln f + \frac{1}{2}\ln^2f\right)X_{t}\lambda dt +\left(\ln f + \frac{1}{2}\ln^2f \right)X_{t^-}d\tilde{N}_t + \mathcal{O}(\lambda\ln^3f)
\ee

\noindent Now let us absorb the mean drift of the Poisson process as scaling factors for the growth rate and carrying capacity

\be
dX_t = r \left(1+\frac{\lambda}{r}\ln f + \frac{\lambda}{2r}\ln^2f \right)X_t\left(1+\frac{X_t}{K\left(1-\frac{\lambda}{r}\ln f + \frac{\lambda}{2r}\ln^2f \right)}\right)dt +\left(\ln f + \frac{1}{2}\ln^2f \right)X_{t^-}d\tilde{N}_t + \mathcal{O}(\lambda\ln^3f).
\ee

Now we apply the transformation [19] with $\alpha$ finite and evaluate $r'$ in terms of $r$ and $K'$ in terms of $K$.  This has the effect of canceling all $\lambda\ln f$ terms, as intended.

\be
dX'_t(\alpha) = r \left(1 + \frac{\lambda'}{2r}\ln^2f' \right)X'_t\left(1-\frac{X'_t}{K\left(1 + \frac{\lambda'}{2r}\ln^2f' \right)}\right)dt +\left(\ln f' + \frac{1}{2}\ln^2f' \right)X'_{t^-}d\tilde{N'}_t + \mathcal{O}(\lambda\ln^3f)
\ee

\noindent where primed variables depend on $\alpha$.  Before taking the $\alpha\to\infty$ limit, we can identify $\lambda'\ln^2f'$ as $\sigma^2$, a finite constant independent of $\alpha$,

\be
dX'_t(\alpha) = r \left(1 + \frac{\sigma^2}{2r}\right)X'_t\left(1-\frac{X'_t}{K\left(1 + \frac{\sigma^2}{2r} \right)}\right)dt +\left(\ln f' + \frac{1}{2}\ln^2f' \right)X'_{t^-}d\tilde{N'}_t + \mathcal{O}(\lambda\ln^3f).
\ee

\noindent We can now evaluate the $\alpha\to\infty$ limit, knowing how $\tilde{N}_t$ transforms:  $\ln f'd\tilde{N}_t\to\sigma dB_t$ in law, $\ln^2 f'd\tilde{N}_t\to 0$, resulting in

\be
\lim_{\alpha\to\infty}dX'_t(\alpha) = dX^*_t = r \left(1 + \frac{\sigma^2}{2r}\right)X^*_t\left(1-\frac{X^*_t}{K\left(1 + \frac{\sigma^2}{2r} \right)}\right)dt +\sigma X^*_{t^-}dB_t.
\ee

The limiting process is an LES model with increased growth parameters $r_{LES} = r(1+(2r)^{-1}\sigma^2)$ and $K_{LES} = K(1+(2r)^{-1}\sigma^2)$.  Comparing this model to the transformed LRC model of [31] using the analytic results for the stationary mean equations (A12) and (A17) reveals that the two models do indeed have the same stationary mean.

\section{An alternative mapping that equates stationary means}

The stationary means of the LRC and LES models can also be equated by using the same growth rate and carrying capacities and mapping $\sigma^2 = -2\lambda\ln f$, as is clear from equations (A12) and (A17).  This mapping provides an alternative method of quantitatively comparing the two models, though one that is perhaps less meaningful than the diffusion limit approach.  It can be understood intuitively by plotting the time evolution of the mean population of both models in the presence and absence of extinction (Supplementary Figure 2A).  In the absence of extinction, both models asymptote to the same value.  In the presence of extinction, the LRC model average decays to zero faster than the LES model average, indicating higher extinction risk.  Computed directly, the mean times to extinction for the LRC model are significantly shorter than for the LES model (Supplementary Figure 2B), supporting the conclusions of the diffusion limit-based method.

\twocolumngrid
\bibliography{collapse_theory_refs.bib}{}
\bibliographystyle{unsrt}

\clearpage
\onecolumngrid
{\raggedright
\huge{Supplementary Figure 1}}

\begin{figure*}[h]
\centerline{
	\includegraphics[width =3.25 in]{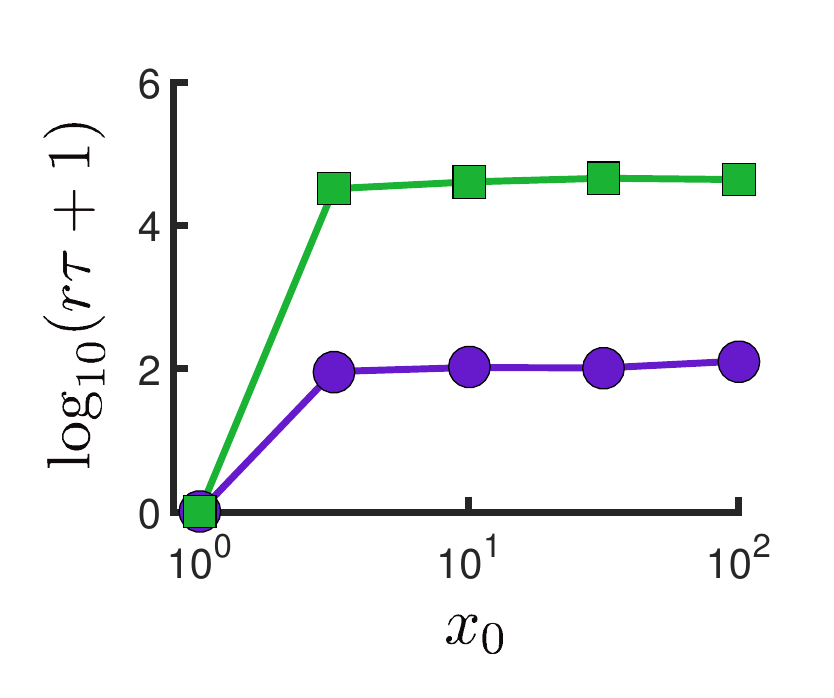}}
	
\end{figure*}

\noindent \textbf{Mean time to extinction is largely independent of initial starting population.}  Mean time to extinction, $\tau$, plotted on a shifted log scale as a function of initial starting population, $x_0$.  Green squares denote the LES model, purple circles denote the LRC model.  The mean extinction time, defined as the first hitting time to $x^* = 1$, starts from 0 but rapidly increases to a value independent of $x_0$.  Parameters:  $r = 1$, $K = 10^4$, $\lambda = .1$, $f = .01$, $N_{trials} = 500$.

\clearpage
{\raggedright
\huge{Supplementary Figure 2}}

\begin{figure*}[h]
\centerline{
	\includegraphics[width =6.5 in]{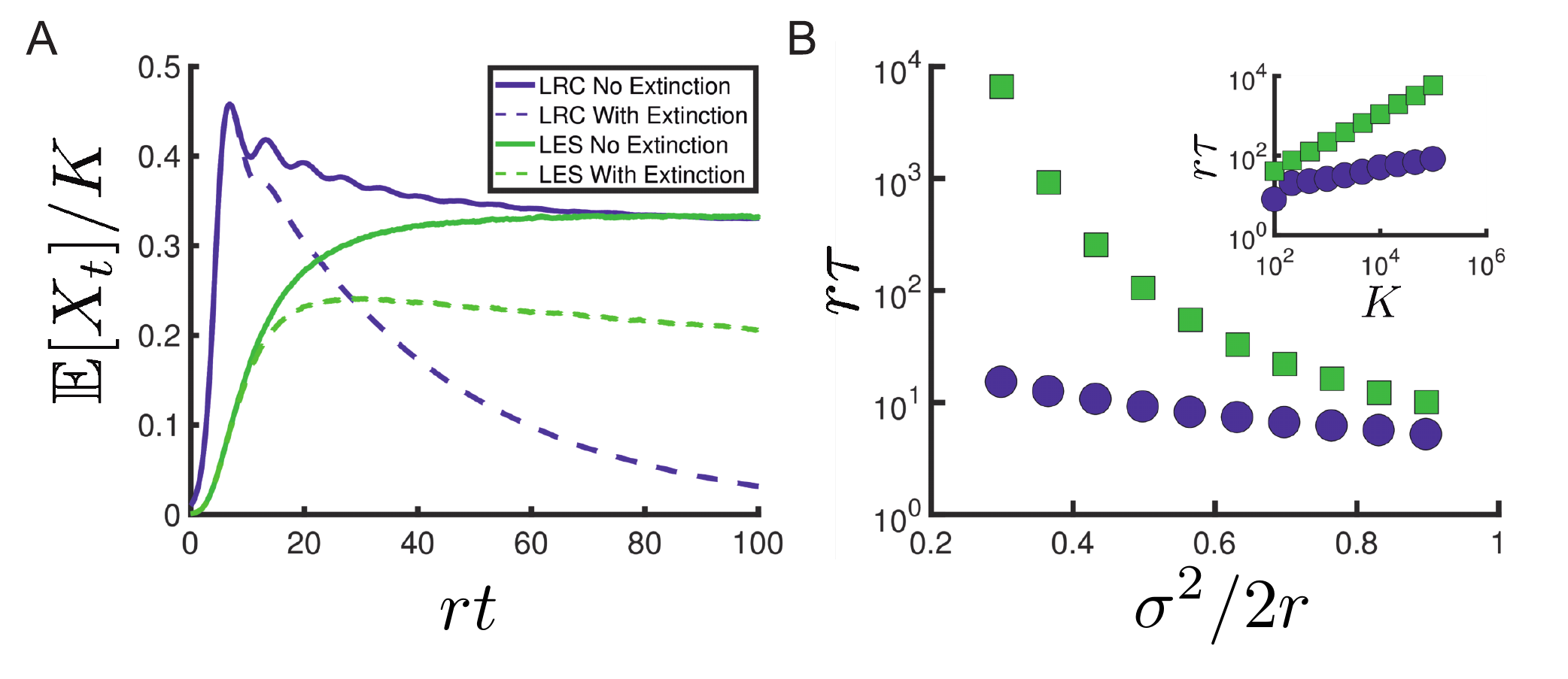}}
\end{figure*}

\noindent \textbf{The LRC model has higher extinction risk than the LES model for equivalent stationary means.} \textbf{A:} Illustration of the mapping.  Numerical results for the average population plotted over time in the LRC (purple) and LES (green) models, showing both the cases of no extinction (dark solid lines) and extinction (light dashed lines) via an absorbing state at $x^* = 1$.  LES model has the same growth rate and carrying capacity as the LRC model and $\sigma$ is determined by $\sigma^2 = -2\lambda\ln f$, such that the two models have equal stationary means (Appendix C).  Parameters:  $r = 1$, $K = 10^4$, $\lambda = .1$, $f = .0012$, $\sigma = 1.16$, $dt = .01$, $N_{trials} = 5\cdot 10^5$.  \textbf{B:}  Mean time to extinction, $\tau$, in units of inverse growth rate, for LRC (purple circles) and LES (green squares) models as a function of noise strength, with $\sigma^2 = -2\lambda\ln f$.   Parameters:  $r = 1, dt = .01, N_{trials} = 5\cdot 10^3$.  For LRC model, $f = .01$ and $\lambda$ was varied from $.065$ to $.195$.  Inset:  Mean time to extinction as a function of carrying capacity.  Parameters:  $r = 1$,  $\lambda = .13$, $f = .01$, $\sigma = 1.09$, $dt = .01$, $N_{trials} = 5\cdot 10^3$. 

\end{document}